\documentclass[fleqn,11pt]{wlscirep}

\usepackage{amsmath,amssymb,graphicx,wrapfig}
\usepackage{textcomp}      % \texttrademark
\usepackage{url}
\usepackage{multirow}
\usepackage{amsmath}
\usepackage{amssymb}

\usepackage{color}
\usepackage{array}
\newcommand{\tb}{\textbf}

\begin{document}

\title{Ultrahigh-Q optomechanical crystals cavities fabricated on a CMOS foundry}

\author[1]{Rodrigo Benevides}
\author[1]{Felipe G. S. Santos}
\author[1]{Gustavo O. Luiz}
\author[1,$\dagger$]{Gustavo S. Wiederhecker}
\author[1,$\dagger$,$\star$]{Thiago P. Mayer Alegre}
\affil[1]{Applied Physics Department, ``Gleb Wataghin'' Physics Institute, University of Campinas, Campinas 13083-859, SP, Brazil}
\affil[$\star$]{alegre@ifi.unicamp.br}
\affil[$\dagger$]{http://nanophoton.ifi.unicamp.br}

\begin{abstract}

Photonic crystals use periodic structures to create forbidden frequency regions for optical wave propagation, that allow for the creation and integration of complex optical functions in small footprint devices. Such strategy has also been successfully applied to confine mechanical waves and to explore their interaction with light in the so-called optomechanical cavities. Because of their challenging design, these cavities are traditionally fabricated using dedicated high-resolution electron-beam lithography tools that are inherently slow, limiting this solution to small-scale applications or research. Here we show how to overcome this problem by using a deep-UV photolithography process to fabricate optomechanical crystals on a commercial CMOS foundry. We show that a careful design of the photonic crystals can withstand the limitations of the photolithography process, producing cavities with measured intrinsic optical quality factors as high as $Q_{i}=(1.21\pm0.02)\times10^{6}$. Optomechanical crystals are also created using phononic crystals to tightly confine the sound waves within the optical cavity that results in a measured vacuum optomechanical coupling rate of $g_{0}=2\pi\times(91\pm4)$~kHz. Efficient sideband cooling and amplification are also demonstrated since these cavities are in the resolved sideband regime. Further improvement in the design and fabrication process suggest that commercial foundry-based optomechanical cavities could be used for quantum ground-state cooling.

\end{abstract}

%\ocis{220.4880,230.1040,230.5750}

\maketitle %%required
%%*********************************
%%*************MAIN************
\newcommand{\nocontentsline}[3]{}
\newcommand{\tocless}[2]{\bgroup\let\addcontentsline=\nocontentsline#1{#2}\egroup}

\section*{Introduction}
Shortly after the first demonstrations of bandgap structures, both at microwave~\cite{Yablonovitch:1991bs} and infrared frequencies~\cite{Lin:1998ix}, photonic crystal structures have emerged as strong candidates for applications in optical and radio-frequency circuits. In particular, the realization of ultra-high optical quality factors~\cite{Sekoguchi:2014er,Song:2005ez} allowed for several applications, such as ultra-small filters~\cite{Debnath:2013db}, fine-tuned demultiplexing devices~\cite{Jugessur:2006ka} and high-sensibility sensors~\cite{Siraji:2015ky}. Additionally, their mode volume smaller than a cubic wavelength~\cite{Painter:1999kv} leads to strong light-matter interaction, where optical nonlinear regimes can be easily accessed \cite{Corcoran:2009dn, Martemyanov:2004fc}. Properly designed photonic crystal cavities (PhCs) may also be used to support localized long-living mechanical modes that can efficiently interact with the optical field~\cite{Chan:2012iy, Alegre:2011vg}. This optomechanical interaction has been used to study the fundamentals of quantum interaction between light and matter~\cite{Chan:2011dy, Riedinger:2016cl} as well as to develop devices for applications~\cite{Metcalfe:2014bh}, such as RF-memories~\cite{Bagheri:2011cra}, accelerometers~\cite{Krause:2012cf}, torque sensors~\cite{Wu:2014jr} and synchronization of mechanical oscillators~\cite{Zhang:2012ks}. However, most of these devices and applications rely on traditional high-resolution electron beam lithography fabrication that is inherently slow, which limits this solution to small-scale applications or research. On the other hand, complementary metal-oxide-semiconductor (CMOS) fabricated structures have proven to be a mature and viable option for photonic~\cite{Schelew2013,Mehta2014,Ooka2015,Xie:it} and optomechanical~\cite{Luan:2014js} structures. 

Here we do a step forward on the CMOS foundry integration approach by designing and fabricating a 2D-optomechanical crystal. We show that it is possible to fabricate a photonic crystal cavity with intrinsic optical quality factors as high as $Q_i=1.21\times10^6$, the largest ever reported based upon optical lithography. Embedding this optical cavity within a phononic shield~\cite{Alegre:2011vg} resulted in optomechanical crystals where several mechanical modes with frequencies around $\Omega_m=2\pi\times2$~GHz are demonstrated having coupling rates as large as $g_0=2\pi\times91$~kHz. Finally, the increase in the mechanical quality factor due to the phononic shield allowed us to demonstrate cooling and amplification of the mechanical mode at low temperature. Our results pave the way for CMOS foundry based optomechanical crystals as building blocks for fundamental studies and applications.

%%%%%%%%% FIGURE %%%%%%%%%%
\begin{figure*}[!t]
\includegraphics[width=\textwidth]{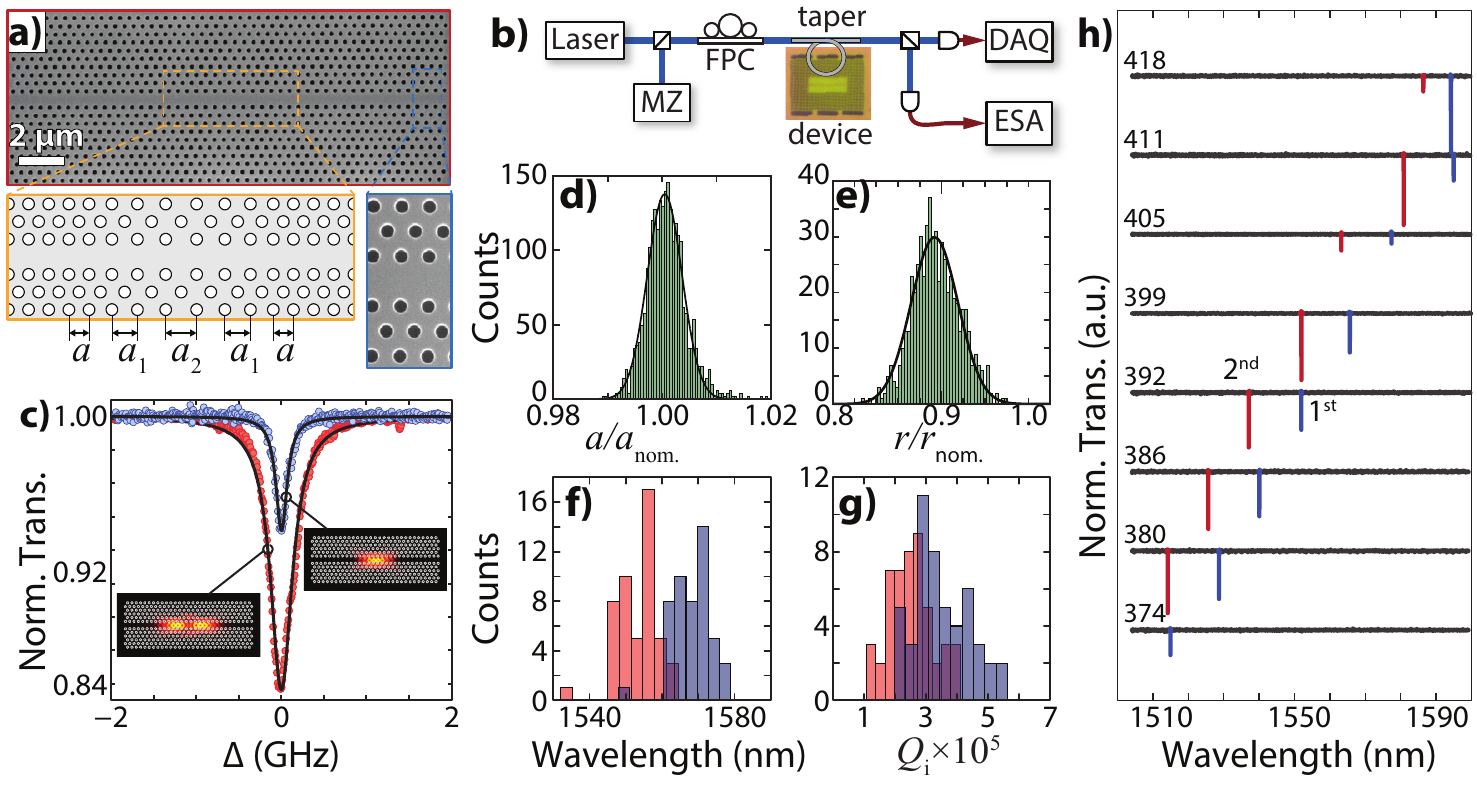}
\caption{\tb{Photonic crystals} \tb{a)} SEM (scanning electron microscope) image of a typical photonic crystal cavity (PhC) with a zoomed image showing the uniformity of the holes. The schematic shows the design of the PhC with the silicon-defect waveguide, in a magnified scale. Around the center of the cavity, the lattice parameter is changed to confine light. The nominal sizes are $a=410$~nm, $a_1=420$~nm and $a_2=430$~nm, hole radius of $r=117$~nm and slab thickness of $t=220$~nm. \tb{b)} Highest optical quality factor observed for the fundamental (blue -- $Q_i=1.21\times 10^{6}$) and second order (red -- $Q_i=6.93\times 10^{5}$) optical modes respectively. The insets show finite element simulations (FEM) of the intensity profile for both modes. \tb{c)} Schematic of the optical setup used for optical and mechanical characterization. A tunable laser wavelength is monitored by a Mach-Zehnder (MZ) interferometer while its radiation is evanescently coupled to the device using a tapered optical fiber, after passing through a fiber polarization controller (FPC). The optical transmission is measured using a slow photodetector (PD) coupled to a data acquisition card (DAQ), while a high-bandwidth PD connected to a electrical spectrum analyzer (ESA), is used to characterize the rapid modulation impinged on the signal by the mechanical oscillation. Distribution of the lattice parameter (\tb{d}) and hole radius size (\tb{e}) based upon high-resolution SEM images. Both values are normalized by their nominal sizes $a_\text{nom.}=410$~nm and $r_\text{nom.}=117$~nm (see Supplementary Material). Central wavelength (\tb{f}) and intrinsic optical quality factor (\tb{g}) measured for the same cavities used to produce the distributions seen on \tb{d)} and \tb{e)}. The blue (red) bars represent the first (second) order optical mode. \tb{h)} Broadband optical transmission spectra for cavities with fixed filling factor ($r/a=0.285$) and lattice parameters from $a=374$~nm to $a=418$~nm. The first and second order optical modes are labelled by the blue and red colors respectively.}
\label{fig:opt-noda}
\end{figure*}
%%%%%%%%% FIGURE %%%%%%%%%%

\section*{Photonic crystal cavities}

One of the main constrains for the realization of optomechanical crystals based on commercial CMOS foundries are the design rules imposed by these facilities. In order to meet the foundry fabrication requirements we have designed a modified version of a photonic crystal cavity based on a previous work~\cite{Song:2005ez}. The device consisted of a hexagonal lattice photonic crystal with a line-defect waveguide and a modification in lattice parameter at the central region to form an optical cavity as shown in Fig.~\ref{fig:opt-noda}\tb{a}. Our approach was to create a deeper optical defect, which means that the change in the lattice parameter was larger than previously reported~\cite{Song:2005ez}. We show that this defect can efficiently confine photons, improving the optical quality factor when compared to other works with photonic crystal cavities fabricated by optical lithography~\cite{Schelew2013,Ooka2015}.

To fully characterize these cavities we did a statistical analysis of their final geometrical and optical properties. The distribution and sizes of the holes of the PhC were determined based on several high-resolution scanning electron microscope (SEM) images from nominally identical cavities as the one shown in Fig.~\ref{fig:opt-noda}\tb{a}. The devices are probed by a tunable laser that is evanescently coupled to the cavity through a tapered optical fiber as shown in Fig.~\ref{fig:opt-noda}\tb{c}. A lorentzian fitted to the DC optical transmission spectra, Fig.~\ref{fig:opt-noda}\tb{b}, yields both the resonant wavelength and linewidth of the cavities.

Since the thickness of the slab is fixed ($t=220$~nm), a given lattice parameter $a$ defines the central position of the photonic bandgap, while the ratio between the hole radius and the lattice size, $r/a$, determines the bandgap width. Figs.~\ref{fig:opt-noda}\tb{d-e} show two histograms for both the lattice parameter $a$ and the hole radius $r$ of these devices extracted from the SEM images, as discussed in Supplementary Material. As we are only interested in the standard deviation of each parameter, as well as the ratio between them ($r/a$), we rescaled each SEM image based on the nominal lattice size. 

A remarkable fidelity of the lattice parameter is seen with a standard deviation smaller than $1$\%. This is corroborated by the measured dispersion of the optical resonant wavelength for both, first (blue) and second order (red) modes, as shown in Fig.~\ref{fig:opt-noda}\tb{f}. On the other hand, the statistics for the hole radius size have two important features. The holes are consistently smaller than its nominal size ($10$\% smaller) and have a larger standard deviation ($5$\%) when compared to the lattice size. The smaller hole sizes could be easily fixed by simply changing the initial hole radius or using a different dose during the photo-lithography process. The standard deviation, however, translates into a disorder in the lattice bands that directly affects the optical confinement and could lead to a smaller optical quality factor. As a result,  the measured intrinsic optical quality factor of these devices, for both the first (blue) and second order (red) optical modes, have a broader distribution than the resonant wavelength as shown in Fig.~\ref{fig:opt-noda}\tb{g}. However they are still centered in a value $Q_i=3\times10^5$, which is larger than previously reported similar devices that are also fabricated by optical lithography~\cite{Xie:it,Ooka2015}. 

Besides the high optical quality factor, another important feature is the ability to design and predict the resonant wavelength of the optical cavities. In order to evaluate this possibility, we have fabricated, in the same chip, several slightly different cavities with lattice parameters and hole radii changing a few percent around a central value ($a=400$~nm $r=114$~nm),  while keeping a fixed filling factor $r/a=0.285$. As a result we would expect roughly a $2.5\%$ change in the cavity's resonant wavelength. The measurements are shown in Fig.~\ref{fig:opt-noda}\tb{h}, where we can observe a steady increase in the resonant wavelength for larger lattice parameter, as expected. 
Conversely fine tuning the radius, while keeping the lattice constant fixed, allows to achieve a desired filling factor. This have a direct impact in the optical quality factors as seen in Fig.~\ref{fig:opt-noda}\tb{b}, where we show the largest measured optical quality factors for the first and second order optical modes. The measured intrinsic optical quality factor for the first order mode is as high as $Q_i=(1.21\pm0.02)\times10^6$, for a cavity with nominal values of $r=121$~nm and $a=405$~nm, showing that a careful design of photonic crystals can withstand the limitations of the photolithography process and produce ultra-high quality optical cavities. 

\section*{Optomechanical crystals}

\begin{figure*}[!t]
\includegraphics[width=\textwidth]{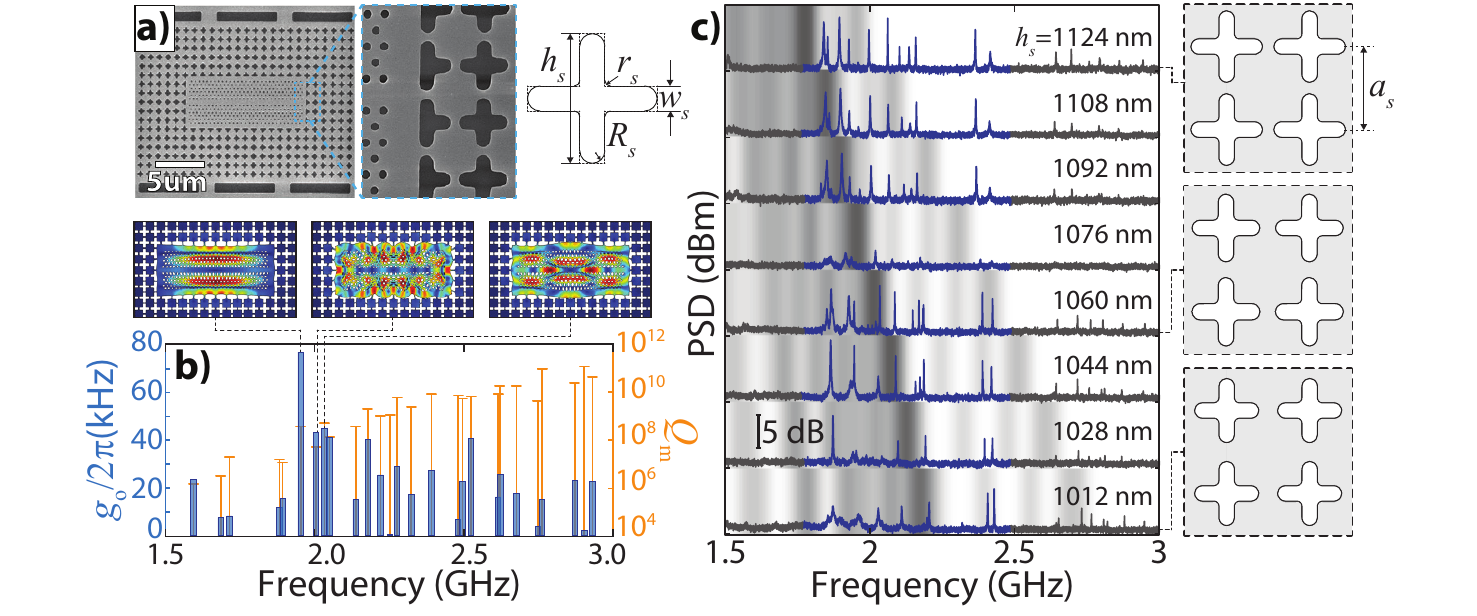}
\caption{\tb{Optomechanical crystals} \tb{a)} SEM images of a fabricated shielded cavity. The inset highlights the rounded shape acquired by the cross holes. The geometrical parameter definitions of the phononic shield are depicted. \tb{b)} Finite element simulations (FEM) for the optomechanical coupling (blue bars, left scale) and mechanical quality factors (yellow lines, right scale) for the PhC surrounded by a shield with values of $a_s=1265$~nm, $h_s=1125$~nm, $w_s=300$~nm.  We observe a logarithmic improvement in quality factors for modes inside the mechanical bandgap, which must be compared to the respective optomechanical coupling rates of each mode. The insets show the mechanical mode displacement profile for the three largest optomechanical coupling rate. The simulation only takes into account mechanical modes symmetric with respect to the central plane of the slab. \tb{c)} Mechanical spectra for eight different mechanical structures, whose crosses lengths ($h_s$) are indicated above the spectra. The shading behind them is related to the density of state for each structure, based on FEM simulations. Darker (ligther) regions correspond to higher (lower) density of states. The simulations parameters were $w_s=297.4$~nm, $R_s=148.7$~nm and $r_s=29.74$~nm and a changing $h_s$, shown in every spectrum.}
\label{fig:noda_shield}
\end{figure*}

To pursue a quasi-2D optomechanical crystal we fabricated a set of PhC embedded into a phononic shield based upon cross-shaped holes (Fig.~\ref{fig:noda_shield}\textbf{a}). The phononic shield enhances both mechanical quality factors and optomechanical coupling rates, thus allowing the measurement of the mechanical modes of the photonic crystal cavity slab~\cite{Alegre:2011vg}. The cross-shaped phononic crystal is designed to have an acoustic bandgap for the in-plane modes with frequencies between $1.5$~GHz and $4.0$~GHz. Performing finite element simulations (FEM) we calculate the optomechanical coupling rate $g_0$ and the radiation limited mechanical quality factor (see Supplemental Material for details) using a shield lattice parameter of $a_s=1265$~nm, $h_s=1125$~nm, $w_s=300$~nm (see Fig.~\ref{fig:noda_shield}\textbf{a} for geometrical definitions) and the nominal values of the PhC cavity shown in Fig.~\ref{fig:opt-noda}\textbf{a}. Following a perturbation theory approach, the optomechanical coupling rate can be calculated as $g_0=-(\omega/2)\langle\vec{E}|\Delta\epsilon|\vec{E}\rangle/\langle \vec{E}|\epsilon|\vec{E}\rangle$, where the inner products ratio between the unperturbed electric field distribution ($\vec{E}$) measures how much the optical mode resonance frequency, $\omega$, varies when the dielectric constant changes from $\epsilon$ to $\epsilon+\Delta\epsilon$ due to the action of the mechanical mode. Strain-induced modifications of the refractive index (photoelastic effect)~\cite{Chan:2012iy} and boundaries deformation~\cite{Johnson:2002jd} are both taken into account for evaluating $\Delta\epsilon$. The phononic shield confines the mechanical modes within the optical photonic crystal cavity slab resulting in simulated optomechanical coupling rates as large as $g_0=2\pi\times(76\pm4)$~kHz (Fig.~\ref{fig:noda_shield}\textbf{b}).

%%%%%%%%% FIGURE %%%%%%%%%%
%%%%%%%%% FIGURE %%%%%%%%%%

As shown in the detail of Fig.~\ref{fig:noda_shield}\textbf{a} however, the resulting fabricated crosses had rounded corners that slightly changed the mechanical bandgaps. To evaluate the role of the phononic shield we designed and tested several nominally identical optical cavities embedded in a cross-shaped phononic crystal with different acoustic shield cross length ($h_s$). The devices are once again probed by a tunable laser that is evanescently coupled to the cavity through a tapered optical fiber resting upon the optical cavity. The typical intrinsic optical quality factor for these cavities is $Q_i= 2.2 \times 10^{5}$ (loaded $\kappa=2\pi\times1.4$~GHz). The phase of the intra-cavity optical field is modulated by the mechanical modes that are excited by thermal fluctuations. The cavity response then converts the phase into amplitude modulation, which is measured using a fast detector and a electrical spectral analyzer (ESA) to determine both the mechanical frequency and linewidth.

In Fig.~\ref{fig:noda_shield}\tb{c} we show the acquired mechanical spectra for samples with varying sizes of the cross length ($h_s$), while keeping the other parameters constant since enlarging $h_s$ also increases the bandgap width size~\cite{SafaviNaeini:2010cu}. In order to take into account the modified geometry we use FEM simulations to calculate the quasi-2D bandgap in-plane phononic bandgap, using the geometries sizes extracted from SEM images. The results are translated into a density of states (DOS) map superimposed to the mechanical spectra, where darker (lighter) shading refers to higher (lower) DOS. The mechanical modes follow the trend of the FEM simulations having higher mechanical quality factors (narrower linewidth) for modes within frequency regions of lower DOS (lighter regions). The measured spectra reveals multiple modes above $2$~GHz that, associated with the narrow optical linewidths, put these modes in the so-called resolved sideband regime ($\kappa\leq\Omega_\text{m}$), where efficient cooling and amplification of the mechanical mode can take place.  

%%%%%%%%% FIGURE %%%%%%%%%%
\begin{figure*}[!t]
\includegraphics[width=\textwidth]{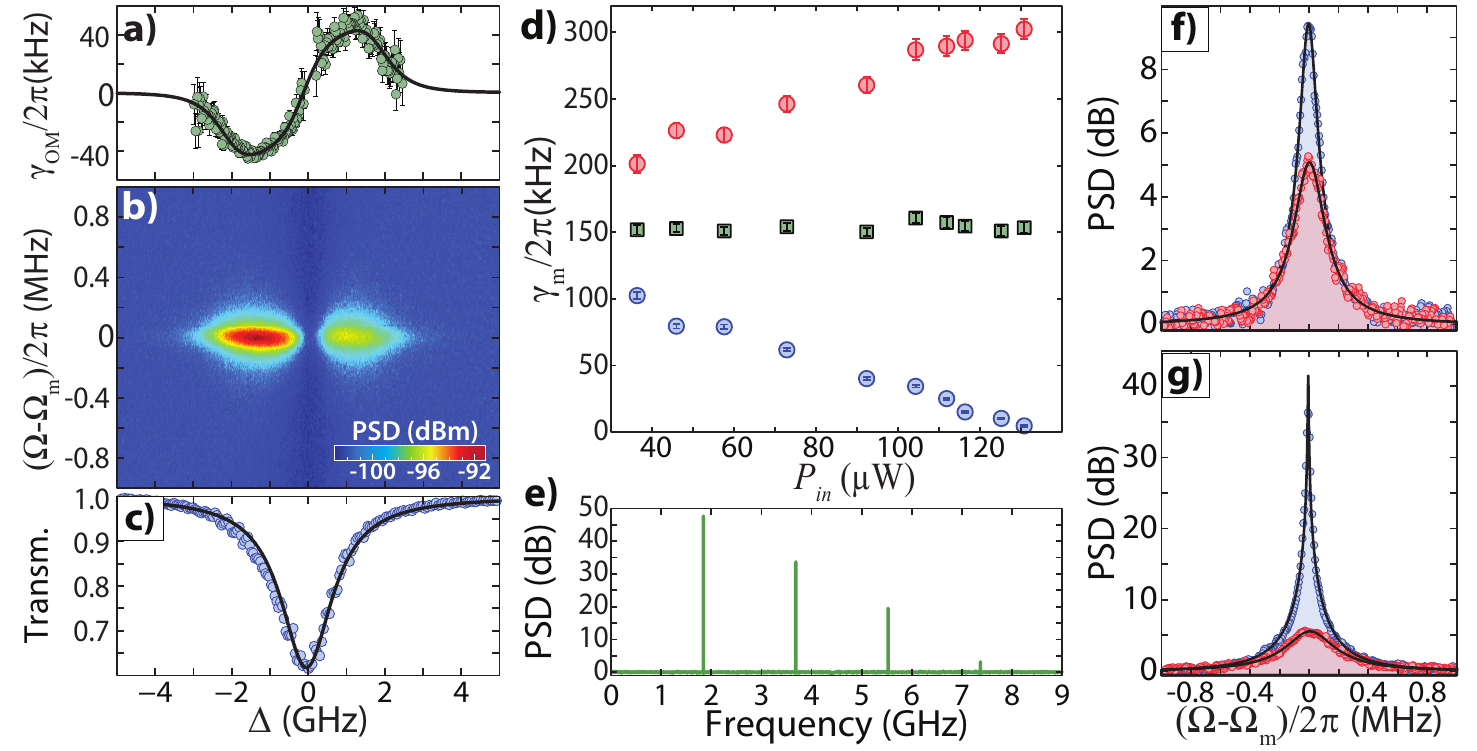}
\caption{\tb{Back-action measurements} \tb{a)} Optomechanical linewidth modification as a function of the laser detuning for the mechanical mode in $\Omega_\text{mec}=2\pi\times1.89$~GHz. The points were obtained by fitting a lorentzian to each mechanical spectra shown in b). The black line indicate the fitting of the theoretical model. \tb{b)} Mapping of the mechanical mode in $\Omega_\text{mec}=2\pi\times1.89$~GHz, as a function of the laser detuning around an optical resonance. \tb{c)} Optical mode used for the sweep in a)-b), with $Q_{i}=1.2\times10^{5}$. Each point in this spectrum corresponds to a mechanical spectrum in b) and a fitting point in a). \tb{d)} The extreme (maxima or minima) effective mechanical linewidth as a function of input power in the cavity. The blue and red dots represent the mechanical linewidth for the laser at the blue ($\Delta<0$) and red sides ($\Delta>0$). The green squares are an average of both measurements which gives the intrinsic mechanical linewidth. \tb{e)} Mechanical spectrum for the cavity in the regime of self-sustained oscillation. The cavity is induced to oscillate in just one mode ($\Omega_\text{mec}=2\pi\times1.89$~GHz), with very high signal and narrower linewidth. A RF frequency comb with frequencies up to $7.5$~GHz is generated. \tb{f)} Mechanical mode for a low input power $P_\text{cav}=36.4\mu W$, for the laser blue and red detuned, with respective resonance colors. Back-action can be already observed. \tb{g)} Increasing the input power up to $P_\text{cav}=146.5\mu W$, we see a $\sim 35$~dB difference between blue and red curves and a remarkable change in linewidth. } 
\label{fig:backaction}
\end{figure*}
%%%%%%%%% FIGURE %%%%%%%%%%

In order to achieve enhanced optomechanical interaction, along with the resolved sideband regime, it is often desirable to have a low mechanical loss in optomechanical cavities. At room temperature, typical mechanical quality factors were $Q_m = 2000$ for modes within the calculated bandgap. Two main loss mechanisms contribute for this modest $Q_m$-factor: the clamping loss associated with imperfections in the phononic shield and phonon-phonon scattering losses (Landau-Rumer) mediated by temperature. While the former is limited by the phononic structure including fabrication imperfections, the latter should be drastically reduce at low temperature.

In order to reveal the influence of the clamping loss term we cooled down the sample using a continuous flow liquid helium cryostat to drastically reduce the phonon-phonon scattering. During the measurements the sample temperature was kept at $25$~K, measured using a calibrated silicon sensor attached to the sample holder. At low temperatures the mechanical losses decreased to $\gamma_i=2\pi\times150$~kHz for a mode at $\Omega_\text{m}=2\pi\times1.89$~GHz, enabling the observation of optomechanical cooling and heating~\cite{Aspelmeyer2014} of the mechanical mode. Using the setup of Fig.~\ref{fig:opt-noda}\tb{c} we scanned the frequency of a tunable laser while simultaneously recording the mechanical power spectrum density (PSD) and the optical transmission. The laser frequency scanning rate ($0.1$~Hz) was intentionally kept much lower that spectrum analyzer acquisition rate ($33$~Hz).

The results are shown in Fig.~\ref{fig:backaction}\tb{a-c} for the lowest input power of $36.8~\mu$W. When the laser is tuned to the blue side of the optical resonance ($\Delta=\omega_l-\omega_0<0$) the mechanical motion is amplified, as indicated by the PSD increase noticeable in the color map of Fig.~\ref{fig:backaction}\tb{b}. This is confirmed by modification of the mechanical linewidth ($\gamma_m=\gamma_i+\gamma_\text{OM}$) as a function of the laser detuning. Fig.~\ref{fig:backaction}\tb{a} shows such modification to the optomechanical damping rate ($\gamma_\text{OM}$), which results in an optomechanical coupling rate of $g_0=2\pi\times(91\pm4)$~kHz and an intrinsic mechanical linewidth of $\gamma_i=2\pi\times(148\pm2)$~kHz. We also explore the relation between the optomechanical damping rate and the optical input power in Fig.~\ref{fig:backaction}\tb{d}, where both blue and red detuned mechanical linewidth are shown. Fig.~\ref{fig:backaction}\tb{f} and \tb{g} are the measured and fitted mechanical spectra for the the lowest and largest optical input powers respectively; the blue and red curves correspond to a laser-cavity detuning matching the mechanical frequency ($\Delta=\pm\Omega_m$).

Above a power threshold ($P_{in}=130~\mu$W in our case) the mechanical mode undergoes a Hopf bifurcation and reaches a self-sustained oscillation regime~\cite{Jenkins:2013dg}. Further increasing the optical input power generates large enough mechanical oscillation to enter a non-linear regime where a RF frequency comb is generated with frequencies up to $7.5$~GHz, as shown in Fig.~\ref{fig:backaction}\tb{e}. Reaching self-sustained oscillation clearly demonstrates that the high optomechanical cooperativity $C=n_\text{cav} g_0^2/\kappa\gamma>1$ (a metric for optomechanical system) limit can be realized in these CMOS-Foundry compatible devices ($n_\text{cav}$ being the intracavity photon number). 

\section*{Conclusion}
In summary we have designed a photonic crystal cavity resilient to the limitations of the photolitography process used in a commercial foundry. As a result we could measure devices with ultra-high Q factors on the order of $10^6$. We also shown how a mechanical crystal can be fabricated in order to create a quasi-2D optomechanical crystal cavity where low mechanical damping loss can be achieved. Finally, low temperature measurements revealed the possibility of using CMOS fabricated structures for efficient back-action measurements. Further improvement in the mechanical shield, along with the presently demonstrated ultra-high optical quality factor, could easily drop the single photon cooperativity by two orders of magnitude thus enabling quantum ground-state cooling in these devices.

%%% SUPPLEMENTAL %%%%

\section*{Methods}

%%%%%%%%%%%%%%%%%%%%%%%%%%%%%%%%%%%%
\subsection*{Fabrication}
\label{sec:fab}
The devices were fabricated through the EpiXfab initiative at IMEC on a silicon-on-insulator wafer (top silicon layer of 220~nm over 2~$\mu$m of buried silicon oxide). The holes and crosses structures were both designed in the same layer. In each optical and optomechanical cavity the holes were referenced to a single hole object in the GDS design. An in-house post-process step was performed to selectively and isotropically remove the buried oxide using a buffered solution of diluted hydrofluoric acid (HF+H$_2$O 1:8) in order to mechanically release the structures. Finally, a piranha (H$_2$SO$_4$+H$_2$O$_2$ 3:1 @ $140~^\text{o}$C) cleaning step to remove any organic residue followed by a HF dip (HF+H$_2$O 1:10) were performed which resulted in an increase of the optical quality factors. 

%%%%%%%%%%%%%%%%%%%%%%%%%%%%%%%%%%%%
\tocless{\section*{Acknowledgements}}
The authors would like to acknowledge CCSNano-UNICAMP for providing the micro-fabrication infrastructure. This research was funded by the Sao Paulo State Research Foundation (FAPESP) (grants 2012/17610-3, 2012/17765-7, 2013/06360-9, 2014/12875-4 and 2016/18308-0), the National Counsel of Technological and Scientific Development (CNPQ - 550504/2012-5, 153044/2013-6), and the Coordination for the Improvement of Higher Education Personnel (CAPES).

%%%%%%%%%%%%%%%%%%%%%%%%%%%%%%%%%%%%
%\tocless{\section*{Author Contributions}}
%R.B., G.S.W. and T.P.M.A. designed the devices. R.B. performed the measurements with support from F.G.S.S., G.O.L. and supervision by T.P.M.A.. R.B., G.S.W. and T.P.M.A. analyzed the measured data. All authors contributed to the writing of the manuscript.

%%%%%%%%%%%%%%%%%%%%%%%%%%%%%%%%%%%%
%\tocless{\section*{Additional Information}}

%\textbf{Supplementary information} accompanies this paper.\\
%\textbf{Competing financial interests:} The authors declare no competing financial interests.

%%*********************************
%%*********BIBLIOGRAPHY************
%\section{References}

%\bibliographystyle{naturemag}

%\bibliographystyle{ieeetr.bst}
\vskip -0.075in
%\tocless{{\small\bibliography{phc_foundry_v2}}}
%\bibliography{phc_foundry}
%\bibliographystyle{unsrt}
\footnotesize
\bibliography{phc_foundry_v2}

\newpage

\begin{center}
\textbf{\LARGE Supplemental Materials: High-Q photonic and optomechanical crystals fabricated on a CMOS foundry}
\end{center}

\vspace{0.5cm}

\noindent \textbf{\Large Rodrigo Benevides, Felipe G. S. Santos, Gustavo de O. Luiz, Gustavo S. Wiederhecker and Thiago P. Mayer Alegre}

\noindent \large Applied Physics Department, ``Gleb Wataghin'' Physics Institute, University of Campinas, Campinas 13083-859, SP, Brazil

%\maketitle %%required
%%*********************************
%%*************MAIN************

\setcounter{figure}{0}
\setcounter{table}{0}
\setcounter{equation}{0}
\setcounter{section}{0}

\renewcommand{\thefigure}{S\arabic{figure}}

%%%%%%%%%%%%%%%%%%%%%%%%%%%%%%%%%%%%
\section*{Lattice parameter and holes size statistics}
\label{sec:statistics}

\begin{figure*}[!h]
%\begin{wrapfigure}{r}{0.35\textwidth}
\centering
\includegraphics[width=0.9\textwidth]{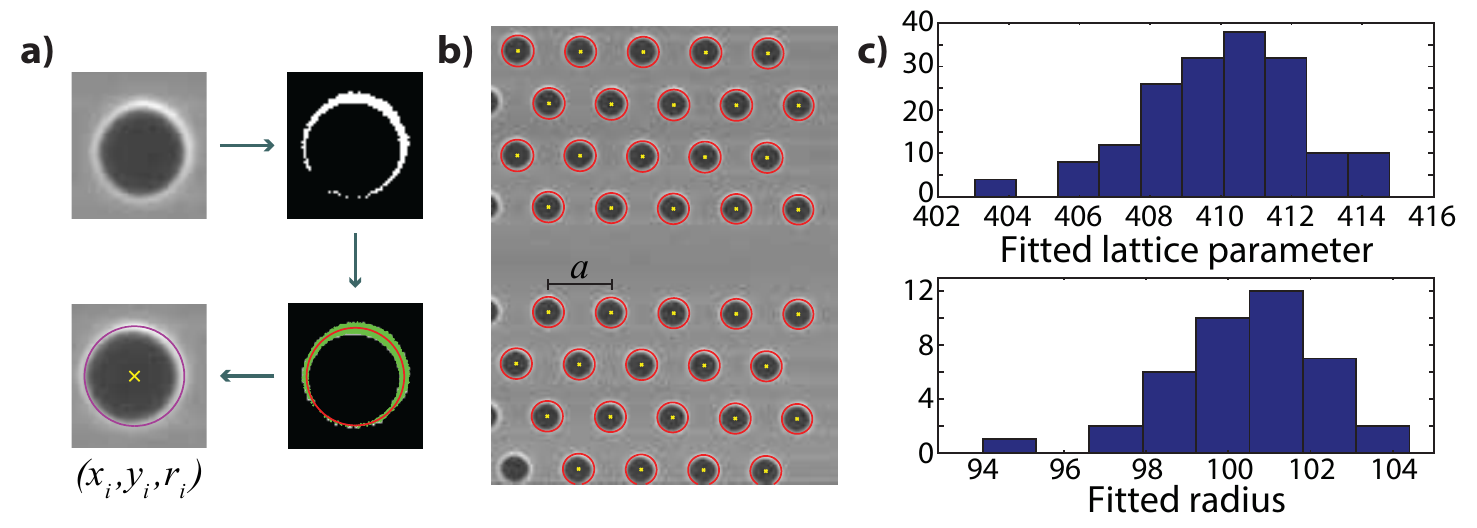}
%\rule{3cm}{2cm}
\caption{\tb{SEM-fitting example} \tb{a)} The process of fitting consists in modifying the image to a black and white picture, based on a given threshold related to the brightness in the original image. A circle is then fitted to the image, providing a center position and a radius. \tb{b)} This process is made for every hole in the picture, resulting in an array of fitted circles. \tb{c)} With the position of the center of the circles we obtain a distribution of distances (top), along with the fitted distribution of radii (bottom).}
\label{fig:SEMfitting}
%\end{wrapfigure}
\end{figure*}

The information about size of holes (radius $r$) and their distances (lattice parameter $a$) were obtained from high resolution ($2048\times 1887$~px) SEM images. The scanned areas were $3.7\times3.4~\mu$m$^2$, which renders the SEM pictures a resolution of $\sim 3.2$~nm$^2$ per pixel. An image processing algorithm was used to fit each hole to a circle as shown in the figure~\ref{fig:SEMfitting}\tb{a}. The fitting results for the holes center coordinates ($x_i$,$y_i$) were used to determine the relative distances between holes $a_{i,j}=\sqrt{(x_{i}-x_{j})^{2}+(y_{i}-y_{j})^{2}}$, which was used as the lattice parameters for each hole pair. The fitting also returned the radius ($r_i$) of each hole.   

The absolute values are measured relative to the averaged measured value for lattice parameters, avoiding possible distortion effect due to astigmatism in the electron microscope. In Fig.~\ref{fig:SEMfitting}\tb{b} we show the resulted fitting for a typical image. All images are taken far from the optical defect region ensuring that the nominal values are retrieved. For this particular picture we have a total of $39$ fitted holes, resulting in $172$ distances. The nominal values for this crystal were $a_\text{nom}=410$~nm and $r_\text{nom}=117$~nm. This process was repeated for each one of the 64 nominally identical cavities in order to obtain the histograms of Figs.~1\tb{d-e} of the main text.

Based on our fitting results, our errors were smaller than 2 pixels, ensuring an uncertainty $<4nm$. All optical measurements were done prior to the SEM images to avoid any contamination from the imaging process.

\section*{Photonic crystal geometry}

\begin{figure*}[!h]
%\begin{wrapfigure}{r}{0.35\textwidth}
\centering
\includegraphics[width=0.9\textwidth]{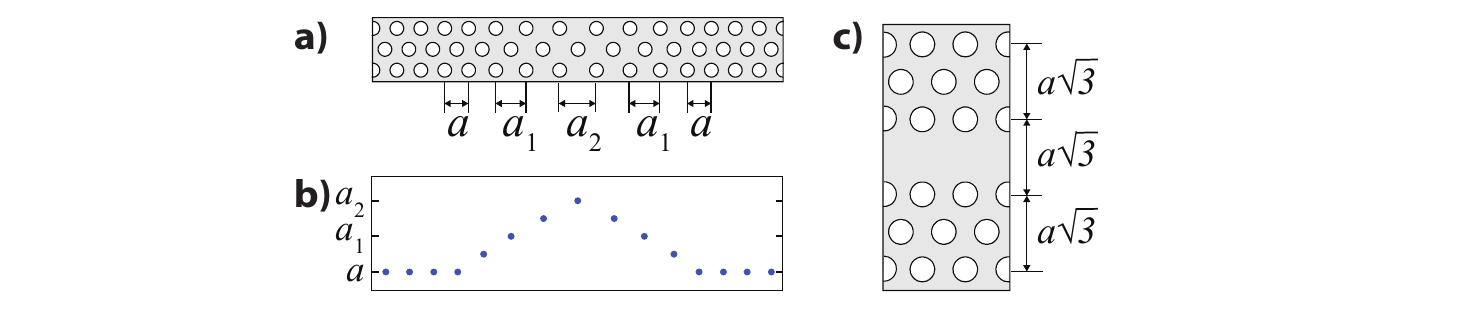}
%\rule{3cm}{2cm}
\caption{\tb{Geometry of the photonic crystal} \tb{a)} A change in the central region of the crystal generates photon confinement. \tb{b)} The change is made smoothly, in an approximately gaussian defect. \tb{c)} A line defect waveguide is created to guide light into the cavity.}
\label{fig:geo}
%\end{wrapfigure}
\end{figure*}

The photonic crystal geometry is based on a 2D-hexagonal lattice, as shown in Fig.~\ref{fig:geo}\textbf{a}. There it is possible to see how the optical cavity is generated in the central region of the slab, using a modification in the lattice parameter. The modification of the holes distances is done in only one direction, such that the effective index of refraction changes smoothly in one axis (parallel to the waveguide) and it remains constant in the other direction. 

In Fig.~\ref{fig:geo}\textbf{b}, we see how the change in the lattice parameter is done, in a continuous way along 10 unit cells, in order to avoid abrupt changes in the effective index of refraction that could lead to photon losses. 

We remove also a line of holes, creating a line defect in the photonic crystal, as shown in Fig.~\ref{fig:geo}\textbf{c}. The bandgap of photonic crystal avoids the propagation of certain wavelengths inside the crystal, creating a waveguide in this line defect. Propagation in the out-of-the-plane direction is avoided for total internal reflection. 

%%%%%%%%%%%%%%%%%%%%%%%%%%%%%%%%%%%%
\section*{Perfectly matched layer parameters in FEM simulations}
\label{sec:pml}
When simulating the optomechanical crystal mechanical radiation loss (Fig.~2\textbf{b} of the main text), we used a cartesian perfectly matched layer (PML) surrounding the phononic crystal device. Inside the PML, the coordinates are stretched in up to three directions according to the polynomial stretching function:
\begin{equation}
f_p(\xi) = s\lambda \xi^p(1-i)
\label{eq:pml}
\end{equation}
where $\xi$ is a dimensionless coordinate varying between 0 (close to the phononic crystal) to 1 (at the end of the computational domain); $\lambda=3.6~\mu$m is the typical mechanical wavelength, $s=3$ a scale factor, $p=1.2$ the curvature parameter. The Q-factor is then obtained from the eigenfrequency $\Omega$ as $Q_m=Re[\Omega]/2Im[\Omega]$.

%%*********************************
%%*************APPENDIX************
%\appendix
%\newcounter{figureS}
\setcounter{figure}{0}
\setcounter{table}{0}
\setcounter{equation}{0}
\setcounter{section}{0}
\renewcommand{\theequation}{S\arabic{equation}}
\renewcommand{\thesection}{S\arabic{section}}
\renewcommand{\thesubsection}{\Alph{subsection}}
\renewcommand{\thesubsubsection}{\roman{subsubsection}}
\renewcommand{\thefigure}{S\arabic{figure}}
\renewcommand{\thetable}{S\arabic{table}}

\end{document}